# On the efficiency of solid-state single photon sources

J.M. Gérard[1], B. Gayral[2] and E. Moreau[3]

[1] CEA-Grenoble/DRFMC/SP2M/Laboratoire « Nanophysique et Semiconducteurs »

17 rue des Martyrs, 38054 Grenoble Cedex 9, France

[2] Schlumberger EPS-SRPC, 1 rue Becquerel, 92142 Clamart, France

[3] CNRS/Laboratoire de Photonique et de Nanostructures,

Route de Nozay, 91460 Marcoussis, France

**Abstract :**

We discuss the efficiency of single photon sources based on a single quasi-monochromatic emitter (such as a semiconductor quantum dot) inserted in a pillar microcavity. We show that their efficiency, which is in principle excellent thanks to the Purcell effect, can be drastically limited by extrinsic cavity losses, such as those related to the scattering by the sidewalls roughness. We present novel design rules for micropillars in view of this application and show that for the well-mastered GaAs/AlAs system more than 70% of the emission can be concentrated into the collimated emission beam associated with the fundamental cavity mode.

The development of efficient solid-state sources of single photons (S4Ps) is a major challenge in the context of quantum communications [1,2] and quantum information processing [3]. S4Ps implement a single emitter able to emit photons one by one, such as a molecule [4], a F-center [5], a semiconductor nanocrystal [6] or quantum dot (QD) [7,8]. Collecting efficiently these single photons is essential for a practical use of the S4P. Considering firstly quantum key distribution, the preparation of a secret key using error correction and privacy amplification is possible only if the probability of detecting a photon in a given time slot is larger than the detectors dark count probability [2]. A low efficiency $\varepsilon$ of the S4P (defined as the probability to emit one photon instead of none for a light pulse) puts therefore a severe constraint on the length of the transmission line, besides its obvious influence on the transmission rate itself. Secondly, a single mode operation of the S4P as well as a good efficiency ($\varepsilon \sim 1$) are required to ensure the operation of quantum optical gates based on single photon interference and the scalability of related quantum computing schemes [3].

The most promising route toward an efficient S4P to date is based on the combination of a quasi-monochromatic emitter and of a three-dimensional (3D) optical microcavity [8-11]; this approach also ensures a single mode operation as recently demonstrated for a S4P based on a single QD in a micropillar [8]. It has been recognized as early as 1991 that small, high Q micropillars can potentially collect a large fraction $\varepsilon$ of the spontaneous emission (SE) of a monochromatic emitter into a single confined mode ($\varepsilon \rightarrow 1$) [12]. This result can be viewed as a consequence of the selective enhancement of the SE into the cavity mode (Purcell effect), which provides a dynamic funnelling of the photons into this specific mode [9-11]. On the basis of time-resolved studies, it has been claimed recently that S4Ps based on a single InAs QD in a GaAs/AlAs pillar microcavity would exhibit $\varepsilon$'s higher than 90% for state-of-the-art structures [9-11], and suggested that $\varepsilon$ could become arbitrarily close to 100% for micropillars displaying higher Purcell factors. Unfortunately, these statements, which are implicitly based on an identification of   and  , are not correct. We show in this Letter that the S4P efficiency   is in general much smaller than the SE coupling factor  , due to several technological imperfections whose role has been overlooked until now. For the well-mastered GaAs/AlAs system however, S4Ps displaying  's as large as 70% could be fabricated with present nanofabrication tools, once an appropriate design of the pillar microcavity is performed. Beyond the present case of single QDs, the present considerations could apply as well for S4Ps based on an appropriate pillar microcavity and some other solid-state emitter, e.g. a semiconductor nanocrystal or a molecule in a solid matrix, which display like single QDs [13,14] a narrow emission line at low temperature [15,16].

Compared to other 3D optical microcavities such as microdisks, microspheres or photonic bandgap microcavities, micropillars are better suited to address the photon collection issue thanks to their directional emission. In a micropillar such as the one shown in figure 1, 3D photon confinement results from the combination of a waveguiding by the dielectric pillar and a reflection by two distributed Bragg reflectors (DBRs) placed on both sides of the cavity layer. The far-field emission diagram of micropillars reflects this confinement geometry. For instance, their fundamental mode is dominated in the far-field by a single sharp lobe centered on the *z* axis; its divergence, of the order of 12° for a 1 µm diameter micropillar, is rather well understood when diffraction at the top end of the pillar is taken into account [17]. For S4Ps based on a single emitter coupled to the fundamental pillar mode, we identify in the following   as the fraction of the SE which goes into this collimated emission beam and can be easily collected.

In practice, only part of the photons emitted into the fundamental confined mode will effectively contribute to this directive emission beam due to additional intrinsic or extrinsic loss mechanisms (   ) . In some cases,   is much smaller than   as will be shown later.

The first, intrinsic, loss mechanism is related to the finite reflectivity of the bottom DBR. For a micropillar built with balanced DBRs, confined photons would have for instance equal probabilities to escape the pillar microcavity from its top or from its bottom. In practice, intrinsic losses can be easily engineered by adjusting the number of periods of both DBRs. For the GaAs/AlAs micropillar shown in figure 1 for instance, the transmission coefficient of the 9-period top DBR (7%) is much larger than the transmission coefficient of the 25 period bottom DBRs (0.1%), in order to establish the transmission through the top mirror as dominant intrinsic photon escape path. Though rather obvious, this first point highlights that and are both qualitatively and quantitatively different.

Extrinsic losses and, first of all, the scattering induced by the roughness of the pillar sidewalls [18-21] are in practice a much more severe subject of concern. This effect is highlighted by studies of the cavity quality factor Q (measured for the fundamental mode of circular micropillars) as a function of the pillar size, as shown in figure 1. We study here two series of GaAs/AlAs pillars fabricated by e-beam lithography and reactive ion etching from two planar microcavities composed by a one-wavelength thick GaAs cavity layer and a 25 period bottom GaAs/AlAs DBR. These cavities differ by the thickness of the top DBR, so that their quality factor $Q_{2D}$ are different; we consider both a « high finesse » sample (15 period top DBR, $Q_{2D}$ =5000) and a « low-finesse » one (9 period top DBR, $Q_{2D}$ =1000). These samples contain a single array of InAs QDs, whose emission is used to probe the photonic modes by photoluminescence [9]. For pillar diameters much larger than /$n$ (where is the emission wavelength in vacuum and $n$ the GaAs refractive index), simple theoretical considerations show that Q should be the same for the planar cavity and for cylindrical micropillars etched from it. This behaviour is indeed observed experimentally for the larger micropillars. However, a clear degradation of Q is observed below a certain critical diameter. This shows that the photon lifetime in the cavity becomes shorter due to the onset of a novel escape path. When micropillars are only partially etched, the diffraction by the finite aperture of the micropillar foot can entail such additional losses [20]. If the whole structure is etched, these additional cavity losses are related to the scattering by sidewall roughness [18-21]. For the smallest pillars (d<1.5µm), extrinsic losses dominate; similar Qs are then obtained independently of $Q_{2D}$, since both series of micropillars (etched using the same process) have similar sidewall roughness. The experimental behaviour for Q is well described by a simple model, which assumes that the scattering probability is proportional to the mode intensity at the surface of the pillar [21], which supports this interpretation.

A second class of extrinsic losses already exist in planar microcavities and will affect as well micropillars. For the well mastered GaAs/AlAs planar microcavities, the highest reported values for Q are 11000 for MBE growth [22] and 11600 for MOCVD growth [21]. Some additional losses (such as residual absorption or scattering by interface roughness) must be included to account theoretically for these values [21,22].

For low-loss microcavities, 1/Q can be written as a sum of terms related to the various loss mechanisms : intrinsic losses due to finite DBR reflectivity ($1/Q_{int}$), extrinsic losses already present in the planar cavity ($1/Q_{ext}$), and furthermore sidewall scattering for micropillars ($1/Q_{scat}$). We can thus write for a micropillar:

$$\frac{1}{Q} = \frac{1}{Q_{2D}} + \frac{1}{Q_{scat}} = \frac{1}{Q_{int}} + \frac{1}{Q_{ext}} + \frac{1}{Q_{scat}} \quad (1)$$

Assuming a negligible transmission for the bottom DBR, we can estimate the proportion of the photons which are emitted into the cavity mode and exit the cavity through the top mirror, thus contributing to the collimated emission beam :

$$\varepsilon = \beta \frac{1/Q_{int}}{1/Q} = \beta \left(1 - \frac{Q}{Q_{scat}} - \frac{Q}{Q_{ext}}\right) \quad (2)$$

We will show now that an analysis of the experimental data shown in figure 2 allows to estimate the relative magnitude of intrinsic and extrinsic losses, as well as , for these micropillars. Firstly, modelling using the standard transfer matrix approach show that $Q_{ext}$ is around 30000, for our high-finesse planar microcavity as well as for other GaAs/AlAs planar microcavities displaying record cavity $Q$s [21,22]. Secondly, $Q_{scat}$ can be easily estimated from the decrease of the micropillar $Q$ with respect to $Q_{2D}$. Simple algebra gives:

$$\varepsilon = \beta \left( \frac{Q}{Q_{2D}} - \frac{Q}{Q_{ext}} \right) \quad (3)$$

As previously discussed [9-11] can be estimated from an analysis of the modification of the emitter's SE rate induced by the cavity. Let us first consider an *ideal* emitter, which is monochromatic on the scale of the mode linewidth, located at the antinode of the fundamental mode of the pillar, in perfect spectral resonance with this mode, and with its electric dipole aligned with the local electric field. As first shown by Purcell [23], the SE rate of this ideal emitter in the cavity mode is $F_p/\tau_0$, where $F_p = \frac{3Q(\lambda/n)^3}{4\pi^2 V}$ is the Purcell factor for this mode, $V$ the effective mode volume and $\tau_0$ the radiative lifetime of the emitter when located in a bulk material of refractive index $n$. Micropillars also support a continuum of "leaky" modes, besides their set of discrete resonant cavity modes [9]. The fraction of the SE which is emitted into the cavity mode, , can be therefore written as :

$$\beta = \frac{F_p}{F_p + \gamma} \quad (4)$$

where $\gamma/\tau_0$ is the SE rate into leaky modes. For InAs QDs, the dipole is randomly oriented in the plane of the cavity ; equation (2) still holds if designs the fraction of the SE which is coupled to the two (polarization-degenerate) fundamental cavity modes of circular micropillars. For microcavities sustaining a non-degenerate fundamental mode, such as elliptical micropillars [8], $F_p$ should be replaced in this expression by $F_p/2$. is experimentally of the order of one, as shown by the study by time-resolved PL of QDs which are out-of-resonance, and only coupled to the leaky modes of a pillar microcavity [9].

An estimate of the Purcell factor $F_p$ is shown in figure 1 for our two series of micropillars. For the high-$Q_0$ series, record $F_p$ values are obtained for diameters in the 1 µm range ($F_p$~30), which highlights their excellent optical quality. As a result, very high values are obtained for the small diameter/high $F_p$ micropillars as shown in figure 2. (Though large, does not reach one in the small pillar diameter limit since $F_p$ displays a maximum around 1 µm.) Quite remarkably, these large values for are obtained in spite of the strong degradation of $Q$ induced by the sidewall scattering. By contrast, this effect plays a much more important role for . Since intrinsic cavity losses are very weak (large $Q_{2D}$), additional losses due to sidewall scattering become dominant for diameters below 3 µm, so that it becomes very detrimental to reduce further the pillar diameter beyond this limit. In the present case, the maximum value for (70%) is obtained around d=3 µm for such state-of-the-art, high $Q$ micropillars.

Until now most efforts on QDs in pillar microcavities have been concentrated on a maximisation of the magnitude of the Purcell effect (and therefore of ). It is clear that this approach is far from being optimal as far as the S4P efficiency is concerned, for two reasons. Firstly, it tends to favor low-$V$ cavities, for which losses due to sidewall scattering are larger. Secondly, high-finesse planar cavities are also preferred, since a large $Q_{2d}$ helps reducing total losses ; this again increases the relative role of scattering losses and absorption losses compared to intrinsic (mirror) losses. displays a maximum which results from a trade-off between the favorable reduction of the cavity volume and the detrimental degradation of $Q$, which are both entailed by a decrease of the pillar diameter. The optimum diameter corresponds roughly to the onset of $Q$'s degradation due to sidewall scattering.

We present in figure 3 a theoretical estimate of as a function of the pillar diameter for various values of the intrinsic cavity losses (or $Q_{2D}$). We assume constant absorption losses ($Q_{ext}$ =30000) and a constant sidewall roughness, similar to the one observed on our two series of micropillars. Note that these assumptions correspond to the best-published results not only for the GaAs/AlAs system, but also for all semiconductor pillar microcavities. For all cases, $\eta$ exhibits qualitatively the same dependence as a function of $d$, but the optimum $d$ depends strongly on $Q_{2d}$; it increases as expected as a function of $Q_{2d}$, since high-finesse microcavities are more sensitive to additional losses. The highest value of $\eta$ (0.73) is obtained for $Q_{2d}$ ~2000 and d~2 µm. Quite remarkably, the maximum value of $\eta$ depends rather weakly on $Q_{2d}$ and remains above 0.7 when $Q_{2d}$ is in the 500 to 5000 range. This behaviour allows to choose (if necessary) the pillar diameter without compromising the collection efficiency $\eta$ of the S4P. This is very interesting in practice, since a reduction of the pillar diameter is the most common approach for reducing the number of emitters present in the microcavity down to unity [8,11].

To conclude, efficient single photon sources can be fabricated by placing a spectrally narrow emitter, such as a QD at low temperature, in a pillar microcavity. When the spatial and spectral matching of the emitter with the fundamental cavity mode is optimal, about 70% of the single photons can be funnelled into the collimated emission beam associated with this mode. Following these design rules for pillars, S4P efficiencies as large as 45% have already been measured in preliminary experiments, for a yet unoptimised QD location [24]. In the context of quantum key distribution, such S4Ps will bring therefore a clear improvement compared to attenuated laser diodes not only through the suppression of two-or-more photon pulses [8], but also through a strong enhancement of $\eta$. By contrast, this limited 70% efficiency might complicate the demonstration of quantum optical gates working with single photons and related quantum computing schemes [3]. Improvements in this field will require additional technological efforts, oriented toward a reduction of the sidewall roughness of etched micropillars. This work is supported by the IST-FET project « S4P ». The authors gratefully acknowledge V. Thierry-Mieg for the growth of the planar microcavities, L. Ferlazzo-Manin for the etching of the micropillars, and G. Patriarche for the TEM.

**Figure captions :**

Figure 1 : Experimental dependence of $Q$ (dots) and theoretical fit (thin lines) as a function of the pillar diameter $d$ for two series of micropillars ( ~1µm). The fitting curves are used to evaluate the dependence of the Purcell factor $F_p$ (thick lines). A transmission electron micrograph obtained on a 1µm diameter pillar is shown in the insert.

Figure 2 : Experimental estimates of       and       for the high-$Q$ series of micropillars (dots). Theoretical curves are calculated using for $Q$ the fitting curve shown in figure 1 a) (  =1µm, $Q_{ext}$ =30000).

Figure 3 : Theoretical estimates of       as a function of the pillar diameter for different values of the intrinsic cavity losses (i.e. of $Q_{2D}$)

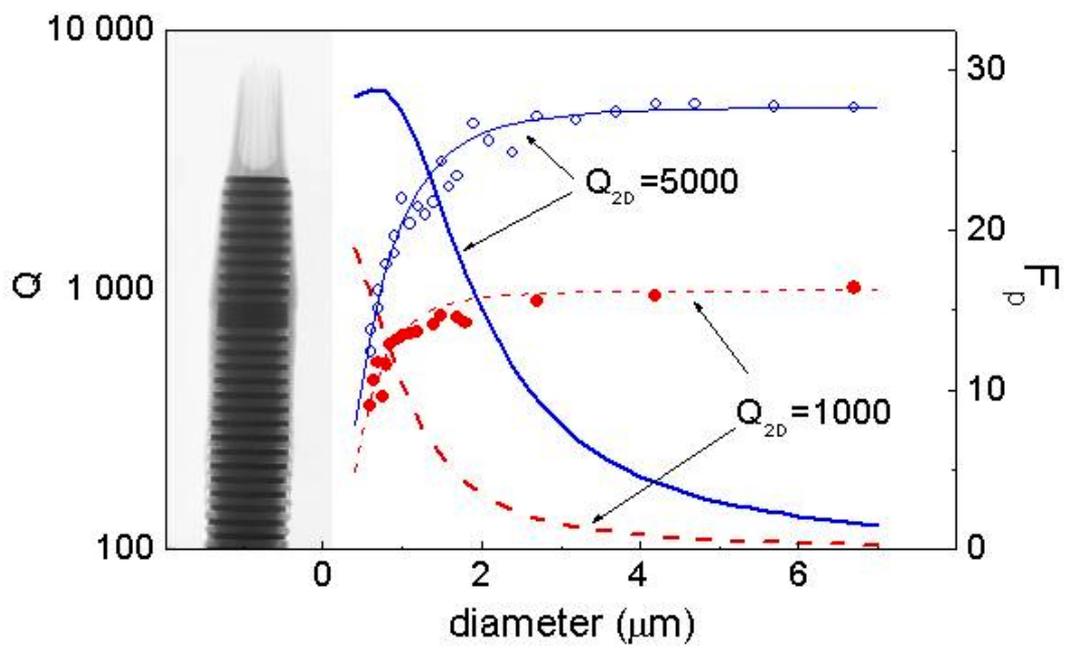

**figure 1**

"On the efficiency of solid-state single-photon sources"  by J.M. Gérard et al

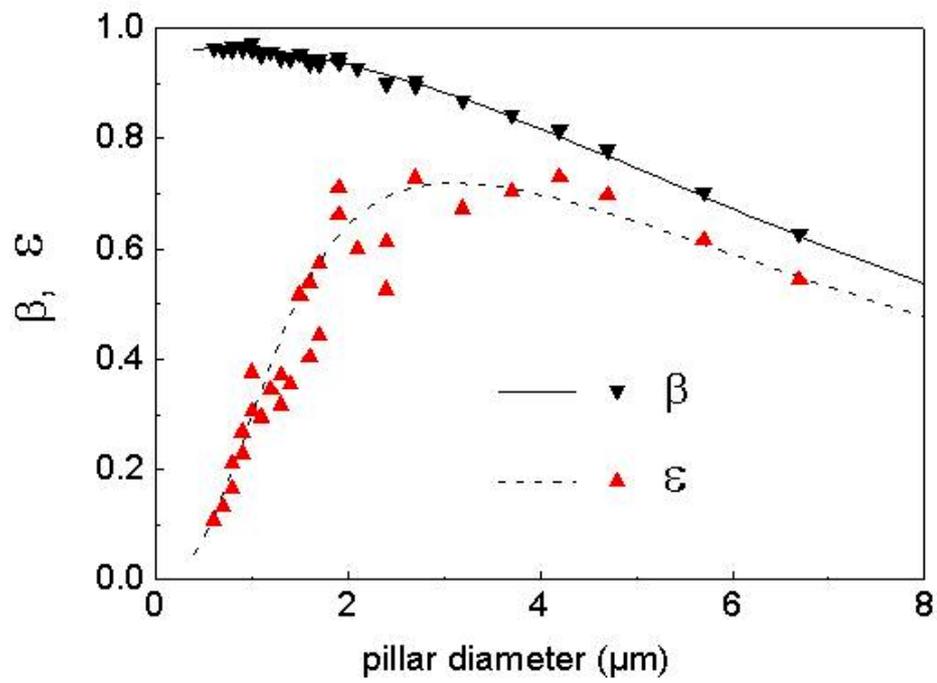

**figure 2**

"On the efficiency of solid-state single-photon sources" by J.M. Gérard et al

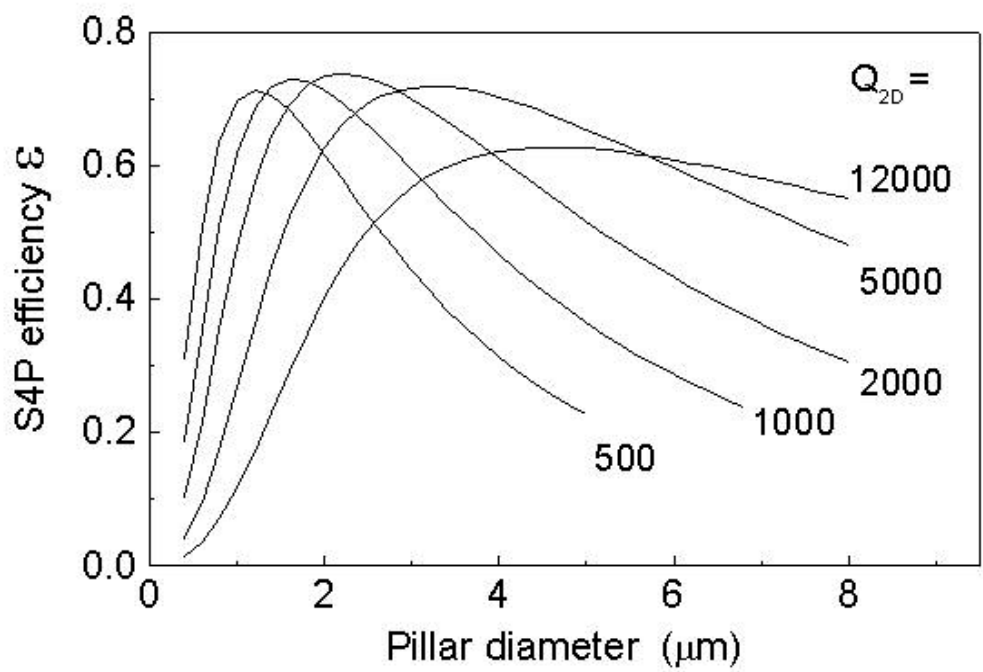

**figure 3**

"On the efficiency of solid-state single-photon sources" by J.M. Gérard et al